\documentstyle[prl,aps,multicol,psfig,array,epsfig,epsf]{revtex}
\draft

\def\sumint{\hbox{$\sum$}\!\!\!\!\!\!\!\int}
\def\square{\vcenter{\vbox{\hrule height.4pt
          \hbox{\vrule width.4pt height8pt
          \kern8pt\vrule width.4pt}\hrule height.4pt}}}

\newcommand{\lsim}{\raisebox{-.1cm}{$\stackrel{<}{\sim}$}}
\newcommand{\beq}{\begin{equation}}
\newcommand{\eeq}{\end{equation}}
\newcommand{\bqa}{\begin{eqnarray}}
\newcommand{\eqa}{\end{eqnarray}}

\begin{document}
\title{The Equation of State for Dense QCD and Quark Stars}

\author{Jens O. Andersen}
\address{Institute for Theoretical Physics, University of Utrecht,
Leuvenlaan 4, 3584 CE Utrecht, The Netherlands}
\author{ 
Michael Strickland}
\address{Physics Department, Duke University, Durham, NC 27701\\
(\today)}

\maketitle

\begin{abstract}
We calculate the equation of state for 
degenerate quark matter to leading order in hard-dense-loop (HDL)
perturbation theory. We solve the Tolman-Oppenheimer-Volkov equations
to obtain the mass-radius relation for dense quark stars.
Both the perturbative QCD and the HDL equations of state have a large variation 
with respect to the renormalization scale for $\mu\,\lsim$ 1 GeV which 
leads to large theoretical uncertainties in the quark star 
mass-radius relation.
\end{abstract}


\begin{multicols}{2}

\section{Introduction}

An understanding of the behavior of quantum chromodynamics (QCD) at 
high density is crucial to describing the physics of compact stars.
This is due to the fact that the nuclear matter within these compact objects
may be sufficiently dense to undergo a phase transition to a deconfined  
or ``quark-matter'' phase.  
Of particular interest is the possibility that stars which have a significant 
quark-matter component could have a dramatically different 
mass-radius relationship than normal neutron stars~\cite{drake}.
However, in order to make definitive statements about the mass-radius 
relationship for such objects one needs a reliable calculation of the equation of
state of high-density QCD.

In recent years high-density QCD has received considerable attention 
due to the possible breaking of color gauge symmetry which gives rise to
color superconductivity. As a consequence of this, QCD has 
a very complicated phase diagram 
which depends on the number and masses of the dynamical quarks~\cite{frankie}.
It would therefore seem that the effect of the superconducting phase
on the equation of state would need to be taken into account; however,
despite the fact that the gaps for color superconductivity are inherently non-perturbative,
their effect on the equation of state for high-density QCD is expected to be 
small~\cite{fraga}.  Therefore, it is possible to use equations of state
that have been derived neglecting them.  

The canonical choice for the equation of state for quark-matter has been
to apply non-ideal bag models with various values for the bag constant~\cite{bags}.
However, others have applied quasiparticle models~\cite{quasi} or 
directly applied the QCD weak-coupling expansion to first or second order 
in the strong coupling constant $\alpha_s$~\cite{fraga}.  The possibility
of applying the weak-coupling expansion is an enticing option since
this expansion can be unambiguously derived from first principles.  However,
since the strong coupling constant is expected to be on the order of one
in the phenomenologically relevant density range, the question of the
convergence of the weak-coupling expansion becomes a very important one.

At high temperatures, for example, the weak-coupling expansion 
for the QCD equation of state  has been carried out to order 
$\alpha_s^{5/2}$~\cite{arnold1,kast,BN}.  Unfortunately, the 
finite-temperature weak-coupling expansion converges very slowly.
In order to improve the convergence of the weak-coupling expansion
calculational frameworks based on hard-thermal-loop
resummation have been proposed~\cite{kpp,EJM1,ejmp,spt,comp1,comp2}. These
approaches attempt to describe finite-temperature QCD in terms of 
weakly-interacting massive quasiparticles by reorganizing the perturbative
expansion around a state which includes hard-thermal-loop quasiparticles
at lowest order.  Detailed studies using these techniques have shown that the 
convergence of the reorganized perturbative series appears to be much better 
than standard resummed perturbation theory\cite{ejmp,spt}.

Motivated by the success of the hard-thermal-loop reorganization of 
perturbation theory we apply the equation of state for dense QCD 
at zero temperature obtained from hard-dense-loop perturbation theory 
(HDLpt) to quark stars.  The goal of doing this is to see if this 
technique can reduce the scale-dependence of the final results and
improve the convergence of the successive approximations to the 
finite-density QCD equation of state.
In this paper we calculate the equation of state of quark-matter
to leading order in HDLpt and use the result to calculate the 
mass-radius relation for non-rotating quark stars.  We also make comparisons 
with the QCD weak-coupling expansion and 
discuss the large theoretical uncertainties in the weak-coupling and HDLpt results.

The paper is organized as follows. In section II, we discuss the weak-coupling
expansion equation of state. In section III, we list the finite-temperature
and density expressions for the leading order HTLpt/HDLpt free energy along 
with the HTL/HDL quark and gluon self-energies.  In section IV, we derive the 
leading order HDLpt equation of state at zero temperature and finite density.
In section V, we use the resulting HDLpt equation of state to determine
the mass-radius relationship of a non-rotating quark star.
Finally, we summarize in section VI.  Necessary integrals are tabulated in
the Appendix.

\section{Weak-coupling expansion}
\label{weaksec}

The zero-temperature, finite-density weak-coupling expansion for the free energy
of an $SU(N_c)$ gauge theory with 
$N_f$ massless quarks has been calculated through order $\alpha_s^2$ by
Freedman and McLerran~\cite{freed}, and by Baluni~\cite{bal} using
the momentum-space subtraction scheme.  In the modified minimal subtraction or 
$\overline{\rm MS}$ scheme the free energy for $N_c=3$ is
\bqa\nonumber
{\cal F}&=&-{N_f\mu^4\over4\pi^2}\Bigg\{1-2\left({\alpha_s\over\pi}\right)
-\left[
A+N_f\log{\alpha_s\over\pi}\right.
\\
&&\hspace{5mm}
\left.+
\left(
11-{2\over3}N_f
\right)\log{\Lambda\over\mu}
\right]\left({\alpha_s\over\pi}\right)^2
\Bigg\}\;,
\label{weak}
\eqa
where $A=A_0-0.536N_f+N_f\log N_f$, $A_0=10.734\pm 0.13$, $\Lambda$
is the renormalization scale, and $\mu$ is the quark chemical potential~\cite{comp2,fraga}.

For the scale dependence of $\alpha_s$ we use the three-loop running
\bqa
\Lambda {\partial \alpha_s \over \partial \Lambda} = -{\beta_0 \over 2 \pi} \alpha_s^2
-{\beta_1 \over 4 \pi^2} \alpha_s^3 -{\beta_2 \over 64 \pi^3} \alpha_s^4 \; ,
\eqa
%
where $\beta_0 = 11 - 2 N_f/3$, $\beta_1 = 51 - 19 N_f/3$, and $\beta_2=2857
-5033 N_f/9 + 325 N_f^2/27$~\cite{qcdreview}. As the boundary condition for 
the integration we fix $\alpha_s=0.1181$ and $N_f=5$ at $\Lambda =M_{Z}= 91.1182$ GeV
and decrease $N_f$ by one as the bottom ($M_{b}$ = 4.0 GeV) and charm ($M_{c}$ = 1.15 GeV) 
quark thresholds are crossed.  Below the charm mass we continue the integration 
with $N_f=3$.  Using this method we obtain $\alpha_s($1 GeV$)=0.4586$.
Note that this method is only approximate since within the $\overline{\rm MS}$ renormalization 
scheme there are non-trivial matching conditions which need to be imposed as each quark 
threshold is crossed; however, the corrections are very small 
so we have ignored them and simply required continuity of $\alpha_s$~\cite{rodrigo}.

\begin{figure}
\epsfysize=7cm
\begin{center}
\centerline{\epsffile{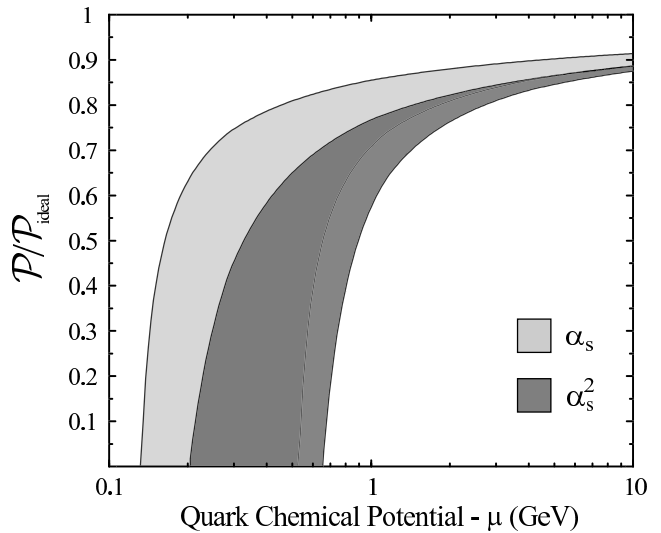}}
\end{center}
\caption{\narrowtext
Perturbative result for the pressure of a 
degenerate quark-gluon plasma as a function of the quark chemical
potential $\mu$ truncated at order $\alpha_s$ and $\alpha_s^2$. 
Bands correspond to variation of the renormalization scale
$\mu \leq \Lambda \leq 4\mu$.
}
\label{fig0}
\end{figure}

In Fig.~\ref{fig0}, we show the pressure (${\cal P}=-{\cal F}$)
 of a degenerate quark-gluon plasma
truncated at order $\alpha_s$ and $\alpha_s^2$. The bands shown are obtained
by varying the renormalization scale $\Lambda$ by a factor of two around 
the central value of $\Lambda=2\mu$.  As can be seen from this figure, 
there is a large theoretical uncertainty in the pressure resulting from the choice
of the scale, particularly for $\mu\,\lsim$ 1 GeV which is the phenomenologically
relevant range.  For example, we see that at NLO the quark chemical potential at which the 
pressure vanishes varies from 200 MeV to 650 MeV.  

Note that the perturbative expansion of the thermodynamical potential is an 
expansion in $\alpha_s$ with coefficients that are polynomial in $\log(\alpha_s)$
and that the logarithms of $\alpha_s$ are due to the effect of plasmons.
To properly assess the convergence of the weak coupling expansion 
of the thermodynamics potential at $T=0$ and $\mu\neq 0$, we need to determine
the order $\alpha_s^3$ contribution. While this would be a very complicated 
calculation, it can be obtained since the entire power series in $\alpha_s$ 
can be calculated using diagrammatic methods~\cite{son,rishke}.
This is in contrast to the high-temperature case where non-perturbative
methods are required at order $\alpha_s^3$ due to infrared divergences associated with
the screening of static magnetic gluons~\cite{linde}.

\section{HDL Free Energy}

The weak-coupling expansion is an expansion about an ideal gas of massless 
particles and the lack of convergence suggests a reorganization of the
perturbative series to improve the convergence and reduce the scale dependence.
One possibility is to use HDLpt which is the analog of 
hard-thermal-loop perturbation theory (HTLpt) 
in the case of large chemical potentials. In HDLpt, one uses effective
propagators and vertices that include plasma effects such as 
propagation of massive quasiparticles, screening, and Landau-damping. 
Thus the expansion point of HDLpt is that of ideal gas of massive
particles. In the case of zero chemical potential and high temperature,
this way of reorganizing the perturbative expansion has dramatically improved
the convergence and reduced the renormalization scale 
dependence~\cite{kpp,EJM1,ejmp,spt}.

At one-loop, HDLpt approximates the high-density phase 
of QCD by a gas of non-interacting massive quasiparticles.  
The one-loop HDL free energy for an $SU(N_c)$ gauge theory with $N_f$ 
massless quarks is

\bqa
{\cal F} &=& (N_c^2-1)\left[(d-1){\cal F}_{T}+{\cal F}_L\right]
\nonumber \\
&&\hspace{2.25cm} + N_c N_f {\cal F}_q + \Delta_0{\cal E}_0\;,
\label{1l}
\eqa
%
where $d = 3-2 \epsilon$ is the number of spatial dimensions
and
${\cal F}_T$ and ${\cal F}_L$ are the contributions to the
free energy from the
transverse and longitudinal 
gluons, respectively\cite{EJM1}.
${\cal F}_q$ is the contribution to the free energy from each color and
flavor of the quarks, and $\Delta_0{\cal E}_0$ is the leading-order 
vacuum-energy counterterm.
The contributions from the transverse and longitudinal
gluons are

\begin{eqnarray}\nonumber
{\cal F}_T & = & {1 \over 2}\sumint_P
\log\left[P^2+\Pi_T(P)\right]\;,
\\
{\cal F}_L & = & {1 \over 2}\sumint_P
\log\left[p^2-\Pi_L(P) \right]
\;,
\label{Fg-def}
\end{eqnarray}
where $\Pi_T(P)$ and $\Pi_L(P)$ are the transverse and
longitudinal HDL gluon self-energies. The quark contribution 
${\cal F}_q$ is 
\bqa
{\cal F}_q=-\log\det\left[P\!\!\!\!/-\Sigma(P)\right]\;,
\eqa
where $\Sigma(P)$ is the quark self-energy. 
The quark contribution can be 
rewritten as~\cite{EJM1,rolf}
\bqa
{\cal F}_q&=&-2\sumint_{\{P\}}\log P^2-2\sumint_{\{P\}}
\log\left[{A_S^2-A_0^2\over P^2}\right]\;.
\label{quark}
\eqa
The functions $\Pi_T(P)$, $\Pi_L(P)$, $A_0(P)$, and $A_S(P)$ are defined by
\bqa
\Pi_T(P)&=&{3m_g^2P^2\over(d-1)p^2}\left[{\cal T}_P-1+{p^2\over P^2}\right]\;,\\
\Pi_L(P)&=&3m^2_g\left[1-{\cal T}_P\right]\;,\\
A_0(P)&=&iP_0-{m_q^2\over iP_0}{\cal T}_P\;,\\ 
A_S(P)&=&p+{m^2_q\over p}\left[1-{\cal T}_P\right]\;,
\eqa
where $m_g$ and $m_q$ are the gluon and quark mass parameters respectively, 
and the function ${\cal T}_P$ is defined by
\bqa
{\cal T}_P&=&
w(\epsilon)\int_0^1dc\;(1-c^2)^{-\epsilon}
{P_0^2\over P_0^2+p^2c^2}
\;,
\eqa
where the function $w(\epsilon)$ is
\bqa
w(\epsilon)&=&{\Gamma({3\over2}-\epsilon)\over\Gamma({3\over2})\Gamma(1-\epsilon)}\;.
\eqa

\section{Zero Temperature and Finite Chemical Potential}
In this section, we calculate the zero-temperature limit of the
one-loop HDL free energy Eq.~(\ref{1l}).
The zero-temperature limit of the gluon contribution was calculated
in Ref.~\cite{EJM1}, while the the zero-temperature
limit of the quark contribution was calculated in Ref.~\cite{rolf}
using three-dimensional expressions for the self-energies. 
Since we are using dimensional regularization in our calculations,
we use $d$-dimensional expressions for them.
To leading order in HDL perturbation theory, 
the difference can be absorbed in a redefinition of the renormalization
scale $\Lambda$. For higher order calculations, it essential to use the
$d$-dimensional expressions for the quark and gluon self-energies.

\subsection{Gluon contribution}
Our original expression for the gluon contribution 
\bqa
{\cal F}_g=(d-1){\cal F}_T+{\cal F}_L
\label{gc}
\eqa
to the 
free energy involves a sum over the discrete Matsubara frequencies 
$\omega_n=2\pi nT$. As $T\rightarrow 0$, the sum approaches an integral
over the continuous energy $\omega$. The only scale in the resulting
integral
is $m_g$ and ${\cal F}_g$ is proportional to $m_g^4$ on dimensional
grounds.

In the zero-temperature limit, the transverse free energy ${\cal F}_T$ becomes
\bqa\nonumber
{\cal F}_{T}&=&
{1\over4\pi}\left({e^{\gamma}\Lambda^2\over4\pi}\right)^{\epsilon}
\int_{-\infty}^{\infty}\;d\omega
\\
&&
\hspace{5mm}
\times
\int_{\bf p}\log\left[
p^2+\omega^2+\Pi_{T}(\omega,p)
\right]\;.
\eqa
Since $\Pi_{T}$ is a function of the combination $\omega/p$ only, 
it is convenient to rescale the energy $\omega\rightarrow p\omega$.
Integrating over the angles of ${\bf p}$ and using the fact that
the integrand is an even function of $\omega$, the integral reduces to
\bqa\nonumber
{\cal F_{T}}&=&{1\over2\pi}
{\Omega_{d}\over{(2\pi})^d}
\left({e^{\gamma}\Lambda^2\over4\pi}\right)^{\epsilon}
\int_0^{\infty}d\omega
\int_0^{\infty}dp\;p^d
\\&&
\hspace{5mm}
\times
\log\left[
(1+\omega^2)p^2+\Pi_{T}(\omega,1)
\right]\;,
\eqa
where $\Omega_d=2\pi^{d/2}/\Gamma(d/2)$ is the angular integral.
The dimensionally regulated integral over $p$ can be evaluated analytically
using Eq.~(\ref{logd}) giving
\bqa\nonumber
{\cal F}_{T}&=&{1\over2\pi}{\Omega_d\over(2\pi)^d}
{\Gamma\left({d+1\over2}\right)\Gamma\left({1-d\over2}\right)\over d+1}
\left({e^{\gamma}\Lambda^2\over4\pi}\right)^{\epsilon}
\\&&
\hspace{5mm}
\times
\int_0^{\infty}d\omega
\left[{\Pi_{T}(\omega,1)\over1+\omega^2}\right]^{(d+1)/2}\;.
\eqa
Expanding around $d=3$ and evaluating 
the resulting integral over $\omega$ numerically, we obtain
\bqa\nonumber
{\cal F}_T&=&
-{9\over8\pi}m_g^4
\left({e^{\gamma}\Lambda^2\over12\pi m_g^2}\right)^{\epsilon}
{\Omega_d\over(2\pi)^d}
\\&&
\hspace{5mm}
\times\left[
\left({1\over\epsilon}+{1\over2}\right){\pi(8\log2-5)\over48}
+0.116815
\right]\;.
\label{ftt0}
\eqa
The longitudinal free energy ${\cal F}_L$ 
can be evaluated in the same manner and 
reads
\bqa\nonumber
{\cal F}_L&=&
-{9\over8\pi}m_g^4
\left({e^{\gamma}\Lambda^2\over12\pi m_g^2}\right)^{\epsilon}
{\Omega_d\over(2\pi)^d}
\\&&
\hspace{5mm}
\times\left[
\left({1\over\epsilon}+{1\over2}\right){\pi(1-\log2)\over3}+
0.320878
\right]\;.
\label{flt0}
\eqa
The gluon contribution ${\cal F}_g$ 
to the free energy is obtained by inserting
Eqs.~(\ref{ftt0}) and~(\ref{flt0}) into Eq.~(\ref{gc}) 
\bqa
{\cal F}_g&=&-(N_c^2-1){9m_g^4\over128\pi^2}\left[{1\over\epsilon}-
2\log{m_g\over\Lambda}+1.24546
\right]\;.
\label{gl}
\eqa
The pole in Eq~(\ref{gl}) agrees with the one found in Ref.~\cite{EJM1},
but the finite term differs since $d$ was set equal to 3 in  the expression
for ${\cal T}_{P}$
\subsection{Quark contribution}
The quark contribution Eq.~(\ref{quark}) can be expanded in a power series
of $m_q^2/\mu^2$. To second order in $m_q^2/\mu^2$, we obtain 
\bqa\nonumber
{\cal F}^{}_q&=&
-2\sumint_{\{P\}}\log P^2-4m_q^2\sumint_{\{P\}}{1\over P^2}
+2m_q^4\sumint_{\{P\}}
\\&&\;
\times\left[{2\over P^4}
-{1\over p^2P^2}+{2\over p^2P^2}{\cal T}_P
-{1\over p^2P_0^2}\left({\cal T}_P\right)^2
\right]\;.
\label{fq}
\eqa
Using the results for the zero-temperature limit of the different sum-integrals
listed in the Appendix, Eq.~(\ref{fq}) reduces to
\bqa
{\cal F}_q&=&-{\mu^4\over12\pi^2}
\left[1-6\left({m_q^2\over\mu^2}\right)
+\left(6-\pi^2\right)\left({m_q^2\over\mu^2}\right)^2\right] \; .
\label{qf}
\eqa
We note that the quark contribution to the free energy is ultraviolet
finite. The coefficient of the $m_q^4$ term is
different from the one found in Refs.~\cite{EJM1,rolf}, 
since $d$ was set equal to 3 in the expression
for ${\cal T}_{P}$.

\subsection{One-loop free energy}
The gluon contribution to the free energy Eq.~(\ref{gl}) is ultraviolet
divergent, while the quark contribution Eq.~(\ref{fq}) is finite.
The ultraviolet divergence is cancelled by the counterterm
that was determined in Ref.~\cite{EJM1}
\bqa
\Delta_0{\cal E}_0&=&{9m_g^4\over128\pi^2\epsilon}\;.
\label{ct}
\eqa
The total free energy is given by the sum of Eqs.~(\ref{gl}),~(\ref{qf}),
and~(\ref{ct}):
\bqa\nonumber
{\cal F}&=&-{N_cN_f\mu^4\over12\pi^2}
\left[1-6\left({m_q^2\over\mu^2}\right)
+\left(6-\pi^2\right)\left({m_q^2\over\mu^2}\right)^2\right]
\\&&
\hspace{5mm}
+(N_c^2-1){9m_g^4\over64\pi^2}\left[\log{m_g\over\Lambda}
-0.622732
\right]
\label{total}
\;.
\eqa
Our leading order result for the thermodynamic functions depends
on the gluon and quark mass parameters, and the renormalization scale
$\Lambda$. These parameters are completely arbitrary in the sense
that the dependence on them will be systematically subtracted out at
higher orders. If higher order calculations were available, 
the masses could be determined by a variational principle giving
rise to a self-consistent gap equation~\cite{kpp,spt}. 
In a one-loop calculation, we have little other
choice than taking the weak-coupling expansion
results for the gluon and quark masses. 
The gluon and quark masses are~\cite{klim+weld}
\bqa
m_g^2&=& {2N_f\over3}{\alpha_s(\Lambda)\over\pi}\mu^2\;,\\
m_q^2&=&{(N_c^2-1)\over4N_c}{\alpha_s(\Lambda)\over\pi}\mu^2\;.
\eqa
In the remainder of the paper we specialize to case $N_c=N_f=3$.
The free energy~(\ref{total}) then reduces to 
\bqa\nonumber
{\cal F}&=&-{3\over4\pi^2}\mu^4\left[
1-4\left({\alpha_s\over\pi}\right)
\right.\\
&& \;
\left.
-3\left(\log{\alpha_s\over\pi}
+2\log{\mu\over\Lambda}+0.0209579
\right)\left({\alpha_s\over\pi}\right)^2
\right]\;.
\label{hdleos}
\eqa
%
Comparing the weak-coupling expansion Eq.~(\ref{weak}) with the leading
order HDL result Eq.~(\ref{total}), we note that the order-$\alpha_s$ term is 
over-included by a factor of two but the $\alpha_s^2 \log\alpha_s$ term is 
included exactly.  A next-to-leading order calculation in HDL perturbation theory
would agree with the weak-coupling expansion at orders $\alpha_s$ and 
$\alpha_s^2 \log\alpha_s$ if we identify the gluon and quark mass parameters 
with their weak-coupling expressions.

\begin{figure}
\epsfysize=7cm
\begin{center}
\centerline{\epsffile{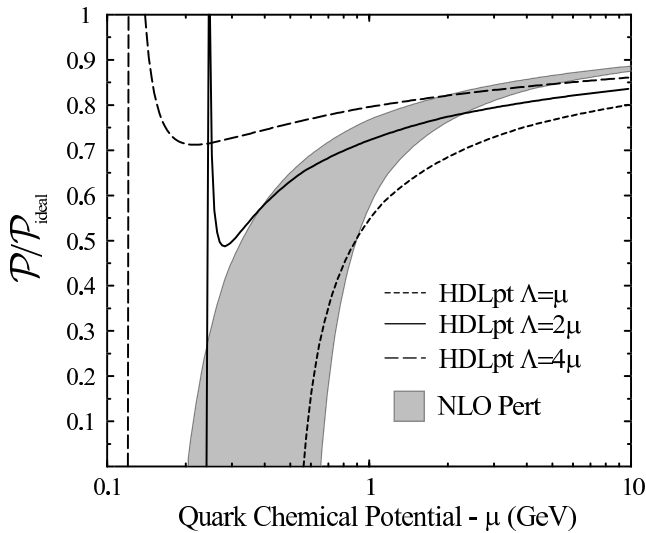}}
\end{center}
\caption{\narrowtext
Leading-order HDLpt result for the pressure of a 
degenerate quark-gluon plasma as a function of chemical potential $\mu$. 
The NLO weak-coupling expansion result is shown as a grey band.
Band corresponds to variation of the renormalization
scale $\mu \leq \Lambda \leq 4 \mu$.  
} 
\label{fig1}
\end{figure}

In Fig.~\ref{fig1}, we show the leading-order HDL result for the free energy
${\cal F}$ normalized to that of an ideal gas. For comparison, we also show the
next-to-leading order prediction from the QCD weak-coupling expansion
as a grey band.  The band is obtained by varying the renormalization scale $\Lambda$
by a factor of two around the central value $\Lambda=2\mu$.
In this figure, we see that the HDLpt results also have a large variation with
respect to the renormalization scale.  Additionally, we see that the results
for both $\Lambda=4\mu$ and $\Lambda=2\mu$ are both unphysical in that they
predict negative quark number densities.  If we add a requirement that the
quark number density be positive we find that this requires 
$\Lambda\,\lsim\,1.6 \mu$.
We will use this as the lower bound for $\Lambda$ in the plots of the mass-radius
relation in the next section.

Note that the requirement that $\Lambda\,\lsim\,1.6 \mu$ may have some
physical basis since the scale of the coupling constant should be related
to the average momentum exchange of two quarks on the Fermi surface.  At zero
temperature the largest momentum exchange possible is $2\mu$ and the 
smallest momentum exchange is of the order of the superconducting gap 
$\phi$.  Therefore, the scale for the coupling constant should be in the range
$\phi\,\lsim\;\Lambda<2\mu$ so that the choice of $\Lambda\sim1.6\mu$
is not unreasonable.

\section{Mass-Radius relationship}

The mass-radius relationship for a non-rotating spherically symmetric 
star is obtained by solving the Tolman-Oppenheimer-Volkov (TOV) equations~\cite{hh}
for the mass $M$ and the pressure (${\cal P}=-{\cal F}$) as a function 
of the radial distance from the center:
\bqa
{d M \over d r} &=& 4 \pi r^2 \tilde{\cal E}(r) \;\\  \nonumber
{d {\cal P} \over d r} &=& - {G \over r^2 c^2}
			\left[\tilde{\cal E}(r)+\tilde{\cal P}(r)\right]
			\left[M(r)+4 \pi r^3 \tilde{\cal P}(r)\right]
\\&& \hspace{5mm}
	\times
			\left[1-{2 G M(r) \over c^2 r}\right]^{-1} \; ,
\eqa
where $G$ is Newton's constant, $c$ is the speed of light, 
$\tilde{\cal E} = {\cal E}/c^2$, and $\tilde{\cal P} = {\cal P}/c^2$.

In this work we will ignore the presence of the nuclear phase of matter
which is expected to undergo a first-order phase transition to
the quark-matter phase.  A more detailed study would include
the effects of the nuclear phase on the mass-radius relationship; however,
our goal here is only to show that both standard perturbation theory and
HDLpt have large theoretical uncertainties related to the renormalization
scale dependence.  The most plausible scenario is that there will not be
``naked'' quark stars, but instead there will be neutron stars with a very
compact quark-matter core and a thick outer layer of normal
nuclear matter.

\begin{figure}
\epsfysize=7cm
\begin{center}
\centerline{\epsffile{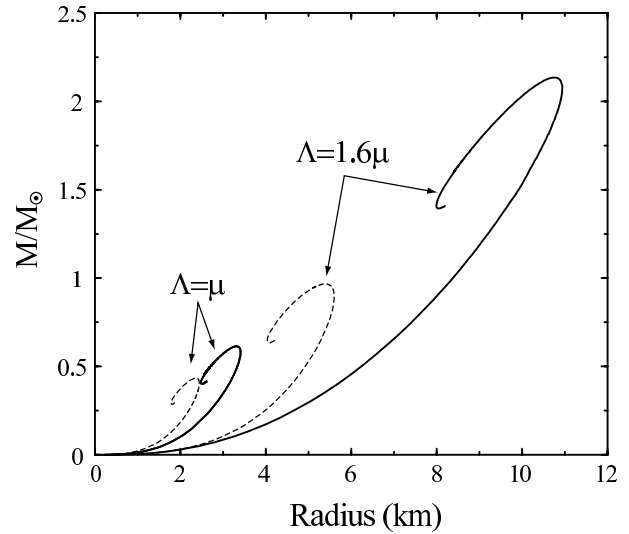}}
\end{center}
\caption[a]{\narrowtext
Mass-radius relation for a quark star with $\Lambda/\mu=1.6$ and $\Lambda/\mu=1$. 
The weak-coupling results for the same choice of renormalization scales are shown as 
dashed lines.  $M_\odot=1.989\times10^{30}$ kg is the mass of our sun.
}
\label{fig3}
\end{figure}

In Fig.~\ref{fig3}, we show the mass-radius relationship obtained by
solving the TOV equations numerically for $\Lambda/\mu=1.6$ and $\Lambda/\mu=1$. 
For comparison, we also show the
QCD weak-coupling expansion results for the same choice of renormalization
scale as dashed lines.  As can be seen from this figure there is a large variation in 
the mass-radius
relationship as the renormalization scale is varied over even this rather limited
range of $\mu \leq \Lambda \leq 1.6\,\mu$.  Using this range, 
we find that using the HDLpt 
equation of state (\ref{hdleos}) that $R_{\rm max} \sim 3.4 - 10.9$ km and 
$M_{\rm max} \sim 0.6 - 2.12 \, M_\odot$.  With this same range we find that using the perturbative equation of state
(\ref{weak}) that $R_{\rm max} \sim 2.4 - 5.6$ km and $M_{\rm max} \sim 0.42 -
0.95 \, M_\odot$.

\section{Discussion}

In this paper, we have calculated the free energy of cold dense 
quark matter to leading order in HDL perturbation theory (HDLpt).
The predictions of HDLpt depend on a renormalization scale
$\Lambda$ that arises both from running of the coupling constant and 
from the renormalization of the additional ultraviolet divergences that
are introduced by the HDLpt reorganization of perturbation theory. 
It is possible to separate these two effects by introducing two 
renormalization scales $\Lambda_3$ and $\Lambda_4$ as done in 
Ref.~\cite{EJM1}, and these scales are associated with the soft scale and hard 
scale, respectively. For simplicity
we have chosen not to distinguish between the two and have simply set 
$\Lambda_3=\Lambda_4$.

We then used the HDLpt equation of state as input to the TOV equations
in order to determine the mass-radius relationship of quark stars.  We
find that the large scale dependence of the HDLpt and weak-coupling
expansion equations of state lead to large theoretical uncertainties
in the quark star mass-radius relationship.  In the case of the weak-coupling
expansion, the expansion only seems to be
under control for $\mu > 5$ GeV.  For HDLpt, the scale dependence is 
larger for all values of $\mu$ and without a next-to-leading order HDLpt
calculation it is not possible to draw conclusions about the convergence
of the series.  
In addition, the choices $\Lambda=4\mu$ and $\Lambda=2\mu$
lead to rather unphysical predictions as can be seen in Fig.~\ref{fig1}.  In order to
eliminate these we were forced to further restrict the range of renormalization
scales considered to $\mu \leq \Lambda \leq 1.6\,\mu$.
The failure of both HDLpt and the weak-coupling expansion to reliably describe the
finite-density QCD equation of state for $\mu$ between
300 MeV and 1 GeV is troubling since this is the range which is important
for determining the mass-radius relationship for a quark star.  

As mentioned in Section \ref{weaksec}, it possible that a computation of 
the order $\alpha_s^3$ contribution to the finite-density QCD equation of state could remove 
some of the theoretical uncertainties resulting from the use of the
weak-coupling expansion result.  Despite the fact that this would be a 
rather difficult task it seems that this calculation is
required in order to draw more firm conclusions about the QCD equation of state.  
However, even if this calculation were available, 
the presence of a non-perturbative 
contribution from a color-superconducting phase of QCD in this range of 
quark chemical potential introduces additional theoretical uncertainties.
Perturbative results extended down to this range of quark chemical potential
give gaps on the order of $\phi \sim$ 30-100 MeV.  Since the gap gives
a relative contribution of the order of $\phi^2/\mu^2$ this could translate
into a relative modification of the equation of state between 1\% and 10\%.

A very challenging problem would be to calculate the next-to-leading order
correction to the free energy in HDLpt.  If the next-to-leading order correction 
turns out to be small for relevant values of the chemical potential, the 
results obtained in the present work can be trusted.  However, it would 
seem more prudent to compute the order $\alpha_s^3$ contribution in the
weak-coupling expansion since the finite-density perturbation series does
not seem to suffer from the same problems (oscillation and lack of convergence)
as the finite-temperature perturbation series.

\section*{acknowledgments}
The authors would like to thank E.~Braaten, M.~Laine, and K.~Rajagopal 
for useful discussions.  J.O.A. was supported by the Stichting voor Fundamenteel 
Onderzoek der Materie (FOM), which is supported by the Nederlandse Organisatie 
voor Wetenschappelijk Onderzoek (NWO).  M.S. was supported by US DOE Grant 
DE-FG02-96ER40945.

\appendix
\renewcommand{\theequation}{\thesection.\arabic{equation}}

\appendix
\renewcommand{\theequation}{\thesection.\arabic{equation}}
\section{}

In the imaginary-time formalism for thermal field theory, 
the 4-momentum $P=(P_0,{\bf p})$ is Euclidean with $P^2=P_0^2+{\bf p}^2$. 
The Euclidean energy $p_0$ has discrete values:
$P_0=2n\pi T$ for bosons and $P_0=(2n+1)\pi T+\mu$ for fermions,
where $n$ is an integer and $\mu$ is the chemical potential. 
Loop diagrams involve sums over $P_0$ and integrals over ${\bf p}$. 
With dimensional regularization, the integral is generalized
to $d = 3-2 \epsilon$ spatial dimensions.
We define the dimensionally regularized sum-integral by
\bqa
  \hbox{$\sum$}\!\!\!\!\!\!\!\int_{P}& \equiv &
  \left(\frac{e^\gamma\Lambda^2}{4\pi}\right)^\epsilon\;
  T\sum_{P_0=2n\pi T}\:\int {d^{3-2\epsilon}p \over (2 \pi)^{3-2\epsilon}}\;,\\ 
  \hbox{$\sum$}\!\!\!\!\!\!\!\int_{\{P\}}&\!\!\!\!\equiv & \!\!
  \left(\frac{e^\gamma\Lambda^2}{4\pi}\right)^\epsilon\;
  T\sum_{P_0=(2n+1)\pi T+\mu}\:\int {d^{3-2\epsilon}p \over (2 \pi)^{3-2\epsilon}}\;,
\label{sumint-def}
\eqa
where $3-2\epsilon$ is the dimension of space and $\Lambda$ is an arbitrary
momentum scale. 
The factor $(e^\gamma/4\pi)^\epsilon$
is introduced so that, after minimal subtraction 
of the poles in $\epsilon$
due to ultraviolet divergences, $\Lambda$ coincides 
with the renormalization
scale of the $\overline{\rm MS}$ renormalization scheme.

\subsection{Simple one-loop sum-integrals}

The simple fermionic sum-integrals required in our calculations are
\bqa
\sumint_{\{P\}} \log P^2 &=& 
{2\mu^4\over3(4\pi)^2}
\; , \\ 
\sumint_{\{P\}} {1\over P^2} &=& 
-{2\mu^2\over(4\pi)^2}\;, \\ 
\sumint_{\{P\}}{1\over(P^2)^2}&=&{1\over(4\pi)^2}
\left({\Lambda\over4\pi\mu}\right)^{2\epsilon}
\left[
{1\over\epsilon}+
2\log(2\pi)
\right]
\;, \\ 
\sumint_{\{P\}}{1\over p^2P^2}&=&{2\over(4\pi)^2}
\left({\Lambda\over4\pi\mu}\right)^{2\epsilon}
\nonumber
\\&&\hspace{5mm}\times
\left[
{1\over\epsilon}+2+
2\log(2\pi)
\right]\;.
\eqa
These sum-integrals can be calculated by standard contour methods.
\subsection{One-loop HDL sum-integrals}
The one-loop fermionic sum-integrals involving the HDL function 
${\cal T}_P$ are
\bqa\nonumber
\sumint_{\{P\}} {1 \over p^2P^2} {\cal T}_P &=&
{2\over (4 \pi)^2} \left({\Lambda\over4\pi\mu}\right)^{2\epsilon}
\\&&
\hspace{-2cm}
\times
\left[ 
\log2
\left({1\over\epsilon}+3\log2+2\log\pi\right)
+{\pi^2\over6}
\right] \;,
\label{t1}
\\ \nonumber
\sumint_{\{P\}} {1 \over p^2 P_0^2}\left({\cal T}_P\right)^2&=&
{4\log2 \over (4\pi)^2} \left({\Lambda\over4\pi\mu}\right)^{2\epsilon}
\\&&
\hspace{-2cm}
\times\left[{1\over\epsilon}+3\log2+2\log\pi\right]
\;.
\label{t2}
\eqa
These sum-integrals can be calculated using the methods developed in 
Ref.~\cite{ejmp}.

\subsection{Integrals}
In order to calculate the zero-temperature limit of ${\cal F}_g$,
we need the following integral
\bqa
\label{logd}
\int_0^{\infty}dp\;p^{\alpha}\log(p^2+m^2)&=&
{\Gamma\left({\alpha+1\over2}\right)\Gamma\left({1-\alpha\over2}\right)
\over\alpha+1}
m^{\alpha+1}\,.
\eqa

\end{multicols}


\begin{thebibliography}{99}
\bibitem{drake} J.J.~Drake, et. al., astro-ph/0204159.
\bibitem{frankie} F. Wilczek, Nucl.Phys. {\bf A663}, 257 (2000). 
\bibitem{fraga}E.S. Fraga, R.D. Pisarski, J.Schaffner-Bielich, Phys. Rev. D {\bf 63}, 121702 (2001).
\bibitem{bags} H.~Satz, Phys. Lett. B {\bf 113}, 245 (1982);
J.~Cleymans, R.V.~Gavai, E.~Suhonen, Phys. Rep. {\bf 130}, 217 (1986).
\bibitem{quasi} A. Peshier, B. K\"ampfer, and G. Soff, Phys. Rev. C 
{\bf 61}, 045203 (2000); hep-ph/0106090.
\bibitem{arnold1}
P. Arnold and C. Zhai, Phys. Rev. D {\bf 50}, 7603 (1994);
	Phys. Rev. D {\bf 51}, 1906 (1995);
\bibitem{kast}
B.~Kastening and C.~Zhai, Phys. Rev. D {\bf 52}, 7232 (1995).

\bibitem{BN}
E.~Braaten and A.~Nieto, Phys. Rev. Lett. {\bf 76}, 1417 (1996); 
       Phys. Rev. D {\bf 53}, 3421 (1996).

\bibitem{kpp}
F. Karsch, A. Patk\'os, and P. Petreczky, 
	Phys. Lett. B {\bf 401}, 69 (1997).
\bibitem{EJM1} 
J.O. Andersen, E. Braaten and M. Strickland, 
	Phys. Rev. Lett. {\bf 83}, 2139 (1999); 
	Phys. Rev. D {\bf 61}, 014017 (2000).
\bibitem{ejmp}J.O Andersen, E. Braaten, E. Petitgirard, and M. Strickland,
hep-ph/0205085.
\bibitem{spt} J. O. Andersen, E. Braaten and M. Strickland,
Phys. Rev. D {\bf 63}, 105008 (2001); J.~O.~Andersen and M.~Strickland,
Phys.\ Rev.\ D {\bf 64}, 105012 (2001).
\bibitem{comp1}
J.-P. Blaizot, E. Iancu, and A. Rebhan, 
	Phys. Rev. Lett. {\bf 83}, 2906 (1999); 
	Phys. Lett. B {\bf 470}, 181 (1999).
\bibitem{comp2}
J.-P. Blaizot, E. Iancu, and A. Rebhan, 
	Phys. Rev. D {\bf 63}, 65003 (2001)
\bibitem{freed}B.A. Freedman and L. McLerran, Phys. Rev. D {\bf 16}, 1130 
(1977); Phys. Rev. D {\bf 16}, 1147 (1977); Phys. Rev. D {\bf 16}, 1168 (1977); 
Phys. Rev. D {\bf 16}, 1108 (1978).
\bibitem{bal}V. Baluni, Phys. Rev. D {\bf 17}, 2092, (1977).
\bibitem{qcdreview}Particle  Data Group, D.E. Groom {\it et al}., 
Eur. J. Phys. C {\bf 15}, 1 (2000).
\bibitem{rodrigo} G.~Rodrigo and A.~Santamaria, Phys. Lett. B {\bf 313}, 441 (1993).
\bibitem{son}D. T. Son, Phys. Rev. D {\bf 59}, 105020 (1999). 
\bibitem{rishke}
R. D. Pisarski D. H. Rischke, Phys. Rev. Lett. {\bf 37}  (1999).
\bibitem{linde} A. D. Linde. Phys. Lett. {\bf B96}, 289 (1980).
\bibitem{rolf}R. Baier and K. Redlich, Phys. Rev. Lett. {\bf 84}, 2100 (2000). 
\bibitem{klim+weld}
V. V. Klimov, Sov. Phys. JETP {\bf 55}, 199 (1982);
H. A. Weldon, Phys. Rev. D {\bf 26}, 1394 (1982).      
\bibitem{hh}H. Heiselberg and M. Hjort-Jensen, 
Phys. Rep. {\bf 328}, 237 (2000).
\end{thebibliography}
\end{document}